# Development of Textured Magnesium Oxide Templates on Amorphous Polymer Surfaces Using Ion-Beam-Assisted-Deposition

Alan J. Elliot*, Ronald N. Vallejo, Rongtao Lu, and Judy Z. Wu

Department of Physics and Astronomy, the University of Kansas, Lawrence, KS 66045

*E-mail address: alane@ku.edu

Biaxially textured MgO templates have been successfully fabricated on several amorphous polymer films including Kapton tapes, polyimide, Poly(methyl methacrylate) (PMMA), and photoresist films using ion-beam-assisted-deposition (IBAD). With a $Y_2O_3$ buffer layer on polymer surfaces, roughening of the polymer surfaces due to preferential ion beam sputtering can be effectively reduced to meet the surface compatibility requirement for IBAD-MgO texturing. In-plane full-width-at-half-maximum (FWHM) of ~10.7º and out-of-plane FWHM ~ 3.5º have been obtained on homoepitaxial MgO films grown on top of the IBAD-MgO template. This method provides a practical route for fabricating epitaxial devices on polymers needed for flexible optoelectronics.

Amorphous or polycrystalline materials, such as glass, ceramics, metals, and polymers, are normally cheap and readily available. To achieve epitaxial growth of technologically important materials on amorphous or polycrystalline surfaces, development of various textured templates using ion-beam-assisted-deposition (IBAD) has been explored.[1-7] Among template materials, magnesium oxide (MgO) has unique advantages in its excellent optical transmittance and a small thickness of ~10 nm needed for the textured IBAD-MgO template, both are ideal for optoelectronic device applications.

On inorganic substrates with hard surfaces, textured IBAD-MgO templates have been obtained with in-plane misalignment of 4° and out-of-plane misalignment of 1.5°.[1, 2] On these templates, epitaxial $YBa_2Cu_3O_7$ superconducting coated conductors[1] and Si(001) films[3] have been fabricated. But IBAD texturing on organic substrates remains challenging although epitaxial growth on polymers is desired for a variety of flexible optoelectronic applications. For example, suspended thin film ferroelectric infrared sensors are typically polycrystalline due to using an amorphous polyimide (PI) substrate as a sacrificial layer.[8] Improved performance is anticipated if they are grown epitaxially on PI since the best ferroelectric effect is along the c-axis,[8] but this remains an unresolved challenge.

The same argument may apply to many other polymers used as resists for photolithography and electron-beam lithography. In addition, PI is a high temperature engineering polymer which exhibits exceptional thermal stability, mechanical toughness, chemical resistance, and a low dielectric constant. PI is commercially available in the form of large-area tape and liquid PI can be spun onto other materials. It has received wide attention as a substrate for roll-to-roll fabrication of flexible and transparent electronics and solar cells.

The difficulties in producing textured IBAD-MgO templates on organic surfaces stem from the incompatibility of the organic surface to the IBAD process. First, the texture quality of the IBAD-MgO template is dictated by the surface morphology of the substrate. An average surface roughness ($R_a$) on the order of 1 nm[8] is generally required to achieve high-quality IBAD-MgO templates. While smooth inorganic surfaces can be readily achieved using standard industrial processes, it is not straightforward on organic surfaces. In addition, the ion-to-atom ratio of the IBAD-MgO process is typically high, in the range of 0.5-1.[2] This means that a direct ion bombardment on the substrate for an extended period during the initial nucleation of the MgO layer is unavoidable, which may cause serious roughening of the surface. On PI films, ion-beam-induced surface roughening has been reported and is attributed to two effects: the physical effect of ion bombardment, and the chemical effect of chain scission due to $O^+$ ion/molecule abstraction of hydrogen followed by the weakening of C-C bonds.[9] Most organic films contain mixed phases of ordered and disordered hydrocarbon chains and the microscopic nonuniformity of the distribution could lead to etch rate nonuniformity across the surface during ion bombardment. Thus, substantial surface roughening seems unavoidable and must be overcome for IBAD-MgO texture development. An amorphous buffer layer may provide a unique solution to this problem. On inorganic substrates, amorphous ceramic layers have been deposited prior to the IBAD process as a seed layer to promote IBAD-MgO nucleation and as a barrier layer to prevent chemical diffusion across the interface between the substrate and the IBAD-MgO layer. In this work, we employed a $Y_2O_3$ buffer layer on polymer surfaces for IBAD-MgO template development. The thickness of the buffer was reduced to a minimum for optimal optical transmittance. We have successfully developed highly textured IBAD-MgO templates on PI films and PI tapes. In addition, two other types of polymer film, poly(methyl methacrylate) (PMMA) and photoresist (PR), were also used to confirm the feasibility of the IBAD texturing process.

Four different types of substrate, 2% PMMA, Shipley 1813 PR, and PI films on Si substrates as well as stand-alone PI tapes (DuPont Kapton HN®), were investigated in this experiment. The PMMA, PR, and PI films were spun onto silicon substrates with a 250-nm-thick thermal oxide layer, and the thicknesses were about 100 nm, 2 μm, and 3.5 μm, respectively. The Kapton tape was 127 μm thick. The fabrication details have been reported elsewhere.[5-7] Briefly, using electron-beam evaporation, a 40 nm film of $Y_2O_3$ was deposited on the polymer surfaces. An $Ar^+$ beam of 750 eV and 115 μA/cm$^2$ was directed at 45° with respect to the substrate surface. The samples were then pre-exposed to the ion beam for different lengths of time. Then, concurrently with the ion beam, MgO was deposited at 1.5 Å/s and a 10 nm textured IBAD-MgO template was deposited. For the PI film samples, a 100 nm MgO film was grown homoepitaxially on the IBAD-MgO template at 300 °C. The template and homoepitaxial films were characterized *in situ* using real-time reflection high-energy electron diffraction (RHEED) to monitor the texture development. $R_a$ of the samples was measured using atomic force microscopy (AFM) at different stages of the IBAD-MgO template development. X-Ray diffraction (XRD) was used to quantify the texture quality.

The $R_a$ of a representative PI film was measured using AFM before and after ion beam exposure for varying times. The original PI film has an $R_a \sim 0.8$ nm in the scan area of 5x5 μm$^2$. After a 2 min ion beam exposure, the $R_a$ increased to 1.6 nm. A 5 min ion beam exposure increased $R_a$ to ~ 3.0 nm. Surfaces of this quality are too rough to obtain well-textured MgO templates. Since Ar and oxygen ions were both present during the exposure, selective sputtering of N and O out of PI may occur simultaneously when the oxygen reacts with hydrogen in phenyl.

Thus, the PI chains would break, generating pits and roughening the surface. Attempts to develop IBAD-MgO templates directly on the PI films without using a buffer layer were not successful using the same IBAD processing parameters as on glass and ceramic substrates.[6]

To protect the PI surface from ion beam exposure roughening while not adding substantial thickness to the IBAD-MgO template to maintain its optical transmittance, a $Y_2O_3$ buffer layer was first evaporated on the $PI/SiO_2/Si$ and then removed partially via $Ar^+$ beam pre-exposure. To optimize pre-exposure time in the $Ar^+$ beam, the surface morphology of this $Y_2O_3/PI/SiO2/Si$ multilayer was measured before and after exposure to the ion beam. The as-grown $Y_2O_3$ buffer layer has an $R_a \sim 1.7$ nm. After a 5 min ion beam exposure, the overall $R_a \sim 2.0$ nm, which is slightly worse than the original $Y_2O_3/PI/Si$ surface morphology. An ion beam exposure of 10 min further increased the surface roughness to $R_a \sim 6.0$ nm. Considering that the average $Ar^+$ beam etch rate of $Y_2O_3$ is $\sim 5.5$ nm/min, the remaining $Y_2O_3$ thickness after 5 min of ion beam exposure is around 5 nm. An additional 5 min of bombardment is mostly on the PI surface.

Since long exposure times in an $Ar^+$ beam roughen the surface of PI, pre-exposure tests were only performed for a short time of 1 min on the other polymer surfaces to compare the effects and also control the roughness within the required range for IBAD-MgO texturing. $R_a$ of the original substrates was measured, as was $R_a$ of the $Y_2O_3$ buffered substrates. The results for PR and PMMA are given in Table I with $SiO_2/Si$ as a reference. All of the surfaces presented in Table I satisfy the requirement of $R_a \sim 1$ nm for IBAD-MgO texturing. After the buffer layer has been exposed to the ion beam, the surface roughens marginally. This means that the buffer layer indeed prevents roughening of the polymer surface under direct exposure of the ion beam. Omitted from Table I is Kapton because, unlike the nearly featureless surfaces of the other polymers, Kapton tape has very smooth regions punctuated by rougher patches. The surface roughness of the Kapton tape actually varies wildly with position and scan area, thus a small area AFM scan cannot reflect the overall surface morphology. Figure 1 shows a photograph of flexed Kapton tape [Fig. 1(a)] and 50×50 μm² AFM scans of the original Kapton [Fig. 1(b)] and the $Y_2O_3$ buffered Kapton after a 1 min $Ar^+$ beam exposure [Fig. 1(c)]. The original Kapton has an $R_a \sim 7.1$ nm over the entire scan area. However, in the 30 μm² subsection, designated by the dashed rectangle, $R_a$ is ~ 2.1 nm. A similar variation is seen in the buffered, pre-exposed Kapton sample where the entire $R_a$ is ~ 8.6 nm, but in the 17 um² dashed rectangle $R_a$ is ~ 1.2 nm. While the bulk Kapton tape does not meet the surface requirement for IBAD-MgO, even after buffering and ion beam pre-exposure, there are sections of the surface that do.

Texture development on these buffered polymer surfaces was monitored using RHEED. The resulting patterns of IBAD-MgO templates on the Kapton tape, PMMA and PR films are shown in Fig. 2, and the one for the PI film is depicted in Fig. 3(a). A 1 min pre-exposure with the $Ar^+$ beam was used on all samples. Clear RHEED patterns indicating cubic MgO (100) texture were obtained on all polymer surfaces. However, superimposed on the RHEED pattern on Kapton tape are several rings indicative of a polycrystalline phase in the film. We suspect that a biaxially textured template was developed on the smooth regions of the Kapton while a polycrystalline film was grown on the rougher regions.

An XRD *θ-2θ* scan, rocking curve, and φ-scan were performed on the PI film samples with an $Ar^+$ beam pre-exposure of 5 min after an additional 100 nm of epitaxial MgO was grown on the IBAD-MgO template. The RHEED pattern clearly improves after homoepitaxial growth. XRD patterns, shown in Figs. 3(c) and 3(d), are used to further evaluate the crystalline quality. Only the MgO (200) peak appears in the XRD *θ-2θ* spectrum shown in Fig. 3(c), which means that the MgO (100) phase is the only crystalline phase in the MgO template. The inset of Fig.

3(c) depicts the MgO (200) rocking curve with a full-width-at-half-maximum (FWHM) ~ 3.5°. The MgO (220) φ-scan indicated in Fig. 3(d) confirms the in-plane FWHM ~ 12°. These results confirm that a highly textured (100) MgO template can be developed on PI.

A summary of the texture quality of a homo-MgO/IBAD-MgO layer fabricated on $Y_2O_3$ buffered PI films with various $Ar^+$ beam pre-exposure times is shown in Table II. Without $Ar^+$ beam pre-exposure, no texture has been observed in either the in-plane or out-of-plane measurements. Although the mechanism is not completely understood, a speculation of "surface activation" at the beginning of the IBAD-MgO process may be necessary to eliminate the contaminants absorbed to the sample surface. The out-of-plane texture quality has been compared using the FWHM of the MgO (200) peak in θ-2θ patterns. The out-of-plane texture quality remains nearly constant with the FWHM of the MgO (200) rocking curves around ~ 3.5°-3.6° for all $Ar^+$ beam pre-exposure times in the range of 2-10 min. But the in-plane texture quality varies. The best in-plane texture was obtained at 2 min $Ar^+$ beam pre-exposure with an FWHM of the MgO (220) φ-scan peak ~10.7°. With increasing $Ar^+$ beam pre-exposure time, the in-plane texture degrades, and at 10 min $Ar^+$ beam pre-exposure no in-plane texture was observed.

In summary, biaxially textured MgO templates have been fabricated on PI, PMMA, and PR films as well as commercially available Kapton tapes using the IBAD texture process. A thin $Y_2O_3$ buffer layer was found to be critical to prevent direct interaction of the ion beam with the polymer surface, which causes serious surface roughening due to the preferential sputtering effects. The successful generation of transparent and biaxially textured MgO templates on polymer surfaces will pave the way for the development of high-performance epitaxial devices for optical and optoelectronic applications.

The authors acknowledge support from NSF EPSCoR for this work. JW was also supported in part by ARO and NSF.

Figures

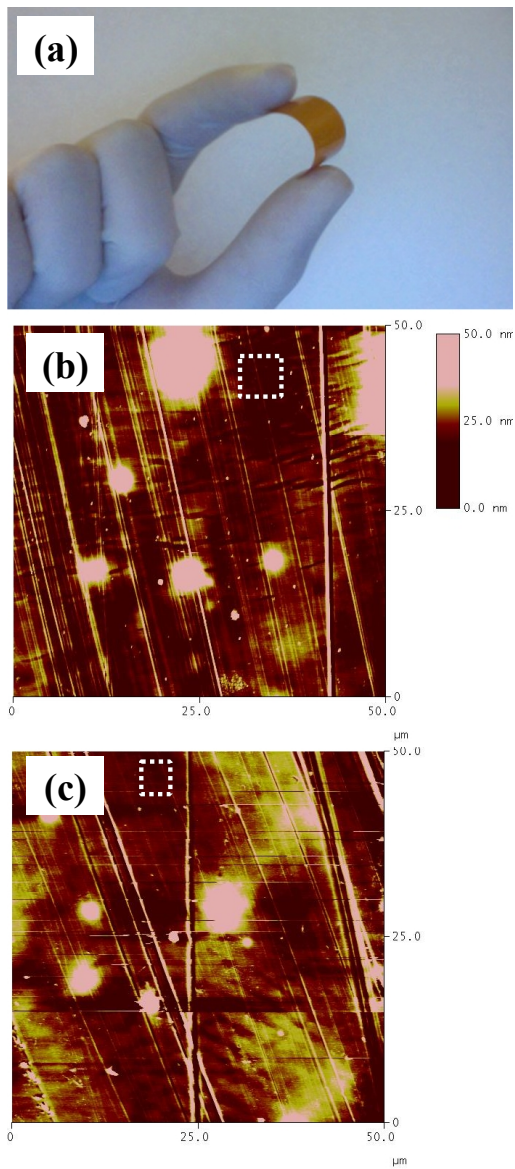

Fig. 1. (a) Photograph of the flexible Kapton tape used in this study. (b) 50x50 µm² AFM scan of original Kapton. $R_a$ over the entire scan area is 7.1 nm, but in the 30 µm² dashed subsection it is only 2.1 nm. (c) 50x50 µm² AFM scan of the $Y_2O_3$ buffered Kapton that has been exposed to the ion beam for 1 min. $R_a$ of the entire area is 8.6 nm, but the roughness of the 17µm² subsection is only 1.2 nm.

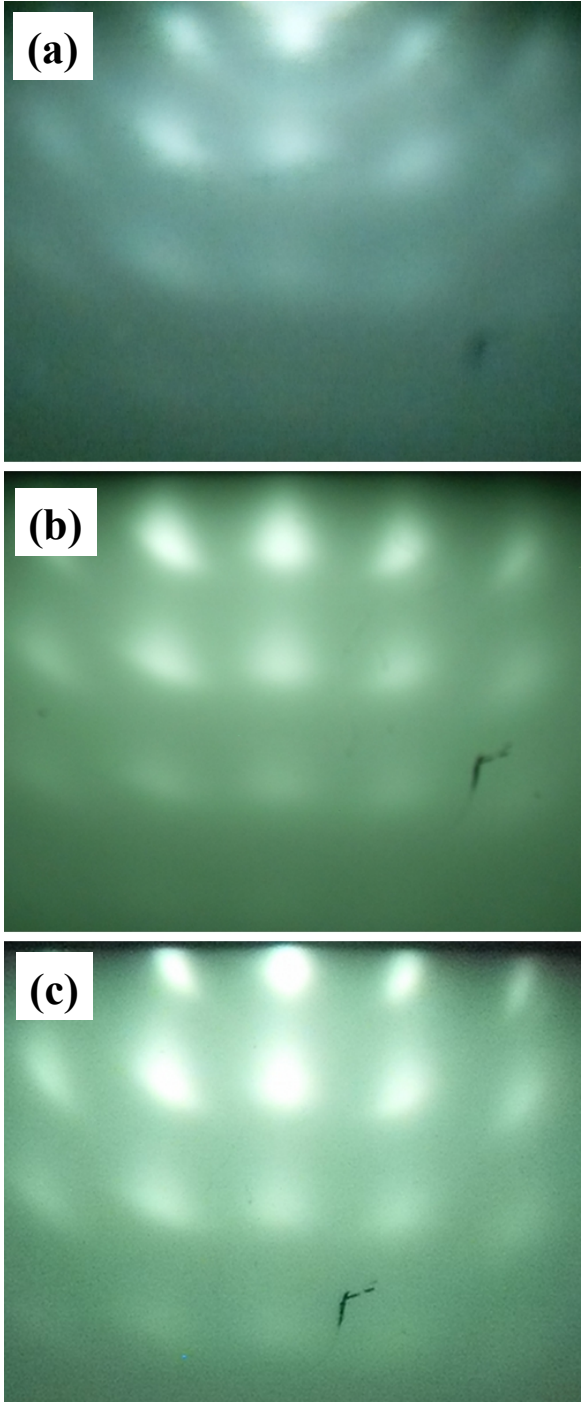

Fig. 2. RHEED patterns for IBAD-MgO templates on (a) Kapton, (b) PMMA, and (c) PR. The substrates were buffered with $Y_2O_3$ then exposed to the ion beam for 1 min.

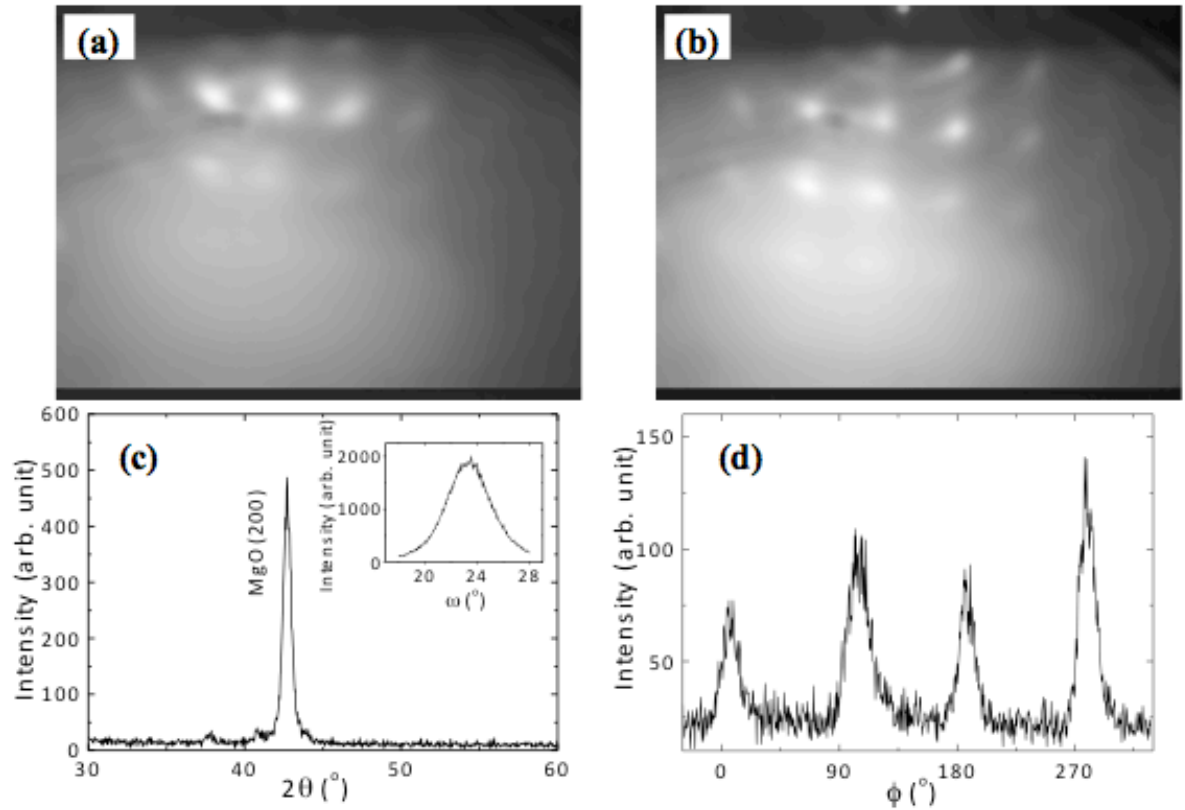

Fig. 3. Texture quality examinations of IBAD-MgO grown on $Y_2O_3$/PI/Si. (a) and (b) are real-time in-situ RHEED patterns of 10-nm-thick IBAD-MgO and 100-nm-thick homoepitaxial MgO layer on top, respectively. (c) and (d) are XRD $\theta$-$2\theta$ scan and MgO(220) $\varphi$-scan patterns for homo-MgO/IBAD-MgO/ $Y_2O_3$/PI/Si, respectively. The inset of (c) shows the MgO(200) rocking curve.

Table I. Surface roughness of original, unexposed substrates and substrates buffered with 30-50 nm $Y_2O_3$ after pre-exposure to the $Ar^+$ beam for 1 min. The scan area was 5×5 $\mu m^2$.

| Surface | Original $R_a$ (nm) | $R_a$ after 1 min Ar+ beam exposure (nm) |
|---|---|---|
| $SiO_2$/Si | 0.3 | 0.8 |
| PR | 0.3 | 0.7 |
| PMMA | 0.4 | 0.7 |

Table II. Texture quality of homoepitaxial MgO/IBAD-MgO vs $Ar^+$ beam pre-exposure time on $Y_2O_3$/PI/Si.

| $Ar^+$ beam pre-exposure time | 0 min | 2 min | 5 min | 10 min |
|---|---|---|---|---|
| **FWHM of MgO (220) φ-scan** | N/A | 10.7° | 12° | N/A |
| **FWHM of MgO (200) rocking curve** | N/A | 3.5° | 3.5° | 3.6° |